\documentclass[epj]{svjour}
\usepackage{graphics}
\usepackage{amssymb}
\begin{document}
\title{New Ideas on SUSY Searches at Future Linear Colliders%
\thanks{Talk presented by S.~Hesselbach}}
\author{S.~Hesselbach\inst{1} 
 \and O.~Kittel\inst{2}
 \and G.~Moortgat-Pick\inst{3}
 \and W.~\"Oller\inst{4}
}                     
%
%
\institute{Institut f\"ur Theoretische Physik, Universit\"at Wien,
  A-1090 Vienna, Austria 
 \and Institut f\"ur Theoretische Physik und Astrophysik,
   Universit\"at W\"urzburg, D-97074 W\"urzburg, Germany and\\
   Instituto de F\'{\i}sica Corpuscular - C.S.I.C.,
   Universitat de Val{\`e}ncia,
   E-46071 Val{\`e}ncia, Spain
 \and IPPP, University of Durham, Durham DH1 3LE, U.K.
 \and Institut f\"ur Hochenergiephysik der \"Osterreichischen Akademie
   der Wissenschaften, A-1050 Vienna, Austria
}
\date{\texttt{DCPT/03/126, HEPHY-PUB 781/03, IFIC-03-47, IPPP/03/63,
UWThPh-2003-39, WUE-ITP-03-021}}
%
\abstract{
Several results obtained within the SUSY group of
the ECFA/DESY linear collider study are presented:
(i) a possibility to determine $\tan\beta$ and the trilinear couplings
$A_f$ via polarisation in sfermion decays, (ii) the impact of
complex MSSM parameters on the third generation sfermion
decays, (iii) determination of CP violation in the complex MSSM via T-odd
asymmetries in neutralino production and decay, and (iv) an analysis of
the chargino and neutralino mass parameters at one-loop level.
} 
\maketitle

\section{Polarisation in sfermion decays: determining $\tan\beta$ and
trilinear couplings}
\label{sec:pol}
The SU(2)$\times$U(1) gaugino mass parameters
$M_2$ and $M_1$, as well as the higgsino mass parameter $\mu$
and $\tan\beta$ for $\tan\beta\le 10$ can be
extracted with high precision
from the chargino/neutralino sector \cite{CKMZ}.
In \cite{Noji} it has been shown that
sfermion production is also
suitable for investigating the properties of neutralinos.
In case of
$\tan\beta\ge 10$ it is appropriate to determine this parameter via
polarisation effects in sfermion decays to fermions plus
neutralinos/charginos in $e^+e^-$ pair production of third-generation
sfermions \cite{staupol},
$e^+e^-\to \tilde{f}_i\bar{\tilde{f}}_j\,,
\tilde{f}_i\to f \tilde{\chi}_k,\quad f=\tau,t,b$.

\begin{figure}
\centering
\resizebox{0.44\textwidth}{!}{%
  \includegraphics{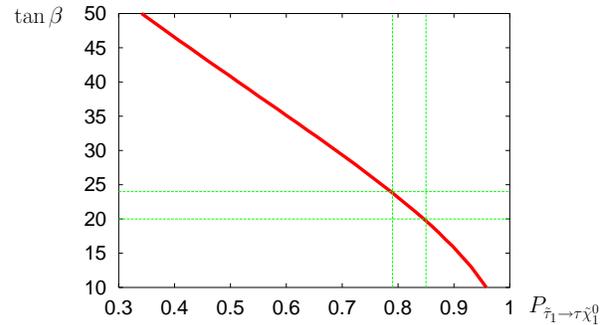}
\setlength{\unitlength}{1cm}
}
  \caption{\label{tau_pol}{
      $\tan\beta$ versus $\tau$ polarisation $P_{\tilde{\tau}_1\to\tau
\tilde{\chi}^0_1}$
      for the reference scenario RP.
      The bands illustrate a measurement of
      $P_{\tau}=0.82\pm0.03$ leading to
      $\tan\beta=22\pm 2$. For details see \cite{staupol}.}}%
\end{figure}

A simulation at one reference scenario RP with $\tan\beta=20$,
inspired by SPS1a \cite{SPS}, is performed in \cite{staupol}.
It is possible to
measure with high precision $m_{\tilde{\tau}_1}=154.8\pm 0.5$~GeV in the
hadronic decay spectra (see e.g. Fig.~7 in the second publication of
\cite{staupol})
as well as the polarisation
$P_{\tilde{\tau}_1\to\tau\tilde{\chi}^0_1}=0.82\pm 0.03$.
The cross section can be measured with an accuracy of
$\delta\sigma/\sigma=3\%$. The use of polarised beams leads to
the unambiguous determination of the mixing angle $\cos 2
\theta_{\tilde{\tau}}=-0.987\pm 0.08$.
The inversion
of the polarisation with respect to its dependence on $\tan\beta$
including the complete gaugino/higgsino mixing leads to the
determination of $\tan\beta=22\pm 2$ (Fig.~\ref{tau_pol}).

An analogous procedure can be applied for $\tilde{t}$ and $\tilde{b}$
production. Since the $t$ polarisation in
the process $\tilde{t}_i\to t \tilde{\chi}^0_k$ depends on
$1/\sin\beta$,
it is only weakly sensitive to large $\tan\beta$.
By contrast, the decay $\tilde{b}_1\to t \tilde{\chi}^{\pm}_1$
can be used indeed to measure $\tan\beta$.
The top polarisation measurement requires the reconstruction of the
$t$ system and of the direction of the primary squark $\tilde{b}_1$ which can
be determined up to a twofold ambiguity and leads to a measurement in the RP
$P_{\tilde{b}_1\to t \tilde{\chi}^0_1}=-0.44\pm -0.10$.

In case that the heavier $\tilde{f}_2$ is accessible, one could
determine the trilinear coupling $A_f$ (for earlier studies see also
\cite{trilinear}),
$A_{f}=[m_{\tilde{f}_1}^2-m_{\tilde{f}_2}^2]/2m_f \sin 2\theta_{\tilde{f}}
+\mu {\tan\beta \choose \cot\beta}$.

A summary of the expected precision of $\tan\beta$ and $A_f$ in our
reference scenario RP is given in Table~\ref{tab_summary} \cite{staupol}.

\begin{table}
\centering
\begin{tabular}{|l|cc|cc|}
\hline  & & & & \\[-2ex]
$\tilde f$ \
              & \multicolumn{2}{|c|}{$\tan\beta$}
              & \multicolumn{2}{|c|}{$A_f$~[GeV]} \\
              & \ ideal \ & \ error \
              & \ ideal \ & \ error \ \\
\hline \hline
 $\tilde{\tau}$& $ 20 $ & $ 2 $   & $ -254 $ & $  - $  \\
 $\tilde{b}$  & $ 20 $ & $ 4.5 $ & $ -773 $ & $ 450 $  \\
 $\tilde{t}$  & $ 20 $ & $ -   $ & $ -510 $ & $  50 $  \\
 \hline
\end{tabular}\caption{{
    Summary of expected precisions of the sfermion mixing parameters
    $\sin 2\theta_{\tilde f}$,
    $\tan\beta $ and the trilinear couplings $A_f$ in reference scenario RP.
    For details on the error estimates see \cite{staupol}.
  \label{tab_summary}}}
\end{table}

\section{Third generation sfermion decays in complex MSSM}
\label{sec:sfdecay}

So far most phenomenological studies on production and decay of
SUSY particles have been performed within the
MSSM with real SUSY parameters.
In \cite{staupapers,squarkpapers} production and decays of the third
generation sfermions in the MSSM with complex parameters $A_\tau$,
$A_t$, $A_b$, $\mu$ and $M_1$ are analyzed.
In a large region of the MSSM parameter space the branching
ratios of $\tilde{\tau}_{1,2}$, $\tilde{\nu}_\tau$, $\tilde{t}_{1,2}$
and $\tilde{b}_{1,2}$ show a strong phase dependence. 
This could have an important impact on the search
for third generation sfermions at a future linear collider and
on the determination of the supersymmetric parameters.

In \cite{staupapers} the effects of the CP phases of $A_\tau$, $\mu$
and $M_1$ on production and decay of $\tilde{\tau}_{1,2}$ and
$\tilde{\nu}_\tau$ are studied. 
The branching ratios of fermionic decays of $\tilde{\tau}_{1}$ and
$\tilde{\nu}_\tau$ show a significant phase dependence for $\tan\beta
\lesssim 10$ whereas it becomes less pronounced for $\tan\beta > 10$.
The branching ratios of the $\tilde{\tau}_{2}$ into Higgs bosons
depend very sensitively on the phases for $\tan\beta \gtrsim 10$.

\begin{figure}
\centering
\resizebox{0.4\textwidth}{!}{%
  \includegraphics{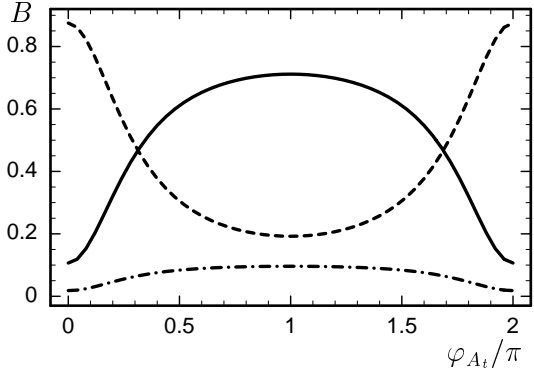}
}
\caption{Branching ratios
$B(\tilde{t}_1 \to \tilde{\chi}^+_1 b)$ (solid),
$B(\tilde{t}_1 \to \tilde{\chi}^0_1 t)$ (dashed) and
$B(\tilde{t}_1 \to \tilde{\chi}^+_2 b)$ (dashdotted)
in a scenario inspired by SPS~1a \cite{SPS}
for $\varphi_{A_b}=\varphi_{\mu}=\varphi_{M_1}=0$, $\tan\beta = 10$ and
\{$m_{\tilde{t}_1}$, $m_{\tilde{t}_2}$, $|A_t|$, $|\mu|$, $M_2$\} 
 = \{379, 575, 466, 352, 193\}~GeV
\cite{squarkpapers}.}
\label{st1decays}
\end{figure}

In \cite{squarkpapers} the impact of the CP phases of $A_t$, $A_b$,
$\mu$ and $M_1$ on the decays of $\tilde{t}_{1,2}$ and
$\tilde{b}_{1,2}$ are analyzed.
The branching ratios of the $\tilde{t}_{1,2}$ show a pronounced phase
dependence in a large region of the MSSM parameter space
(Fig.~\ref{st1decays}).
In the case of $\tilde{b}_i$ decays there can be appreciable
$\varphi_{A_b}$ dependence, if $\tan\beta$ is large and the decays
into Higgs bosons are allowed.

Further the expected accuracy in determining the supersymmetric
parameters was estimated by a global fit of measured masses, branching
ratios and production cross sections. $A_\tau$, $A_t$ and $A_b$
can be expected to be measured with 10\,\%, 2 -- 3\,\% and 50\,\%
accuracy, respectively,
$\tan\beta$ with 1\,\% (2\,\%) accuracy in case
of small (large) $\tan\beta$ and the other parameters with 1\,\% accuracy.

\section{CP violation in MSSM with complex parameters}
\label{sec:cpviol}

The phases $\varphi_{M_1}$ and  $\varphi_{\mu}$
have also impact on the phenomenology of neutralino
production and decay at a future linear $ e^+e^-$
collider and give rise to CP- and T-odd
observables. Such observables, which involve triple products
\cite{triple_products}, may be large, because they
already arise on tree level. In addition, they
also allow the determination of the sign of the phases.
In neutralino  production
(for recent studies see \cite{CKMZ,choi1}):
\begin{equation} \label{production}
   e^+ + e^- \to \tilde{\chi}^0_i + \tilde{\chi}^0_j
\end{equation}
and the subsequent leptonic two-body decay of one of the neutralinos
and of the decay slepton
\begin{equation} \label{decay_2}
   \tilde{\chi}^0_i \to \tilde{\ell} + \ell_1,\quad
   \tilde{\ell} \to \tilde{\chi}^0_1+ \ell_2,\quad \ell_{1,2}=
                e,\mu,\tau,
\end{equation}
the triple product
$ \mathcal T=( {\bf  p}_{e^-} \times
{\bf p}_{\ell_2})\cdot {\bf p}_{\ell_1}$
defines the T-odd asymmetry of the cross section $\sigma$
for the processes (\ref{production}), (\ref{decay_2}):
\begin{equation} \label{Tasymmetry}
{\mathcal A}_{\rm T} = \frac{\sigma({\mathcal T}>0)
                                 -\sigma({\mathcal T}<0)}
                                        {\sigma({\mathcal T}>0)+
                                        \sigma({\mathcal T}<0)}.
\end{equation}

\begin{figure}
\centering
\resizebox{0.4\textwidth}{!}{%
  \includegraphics{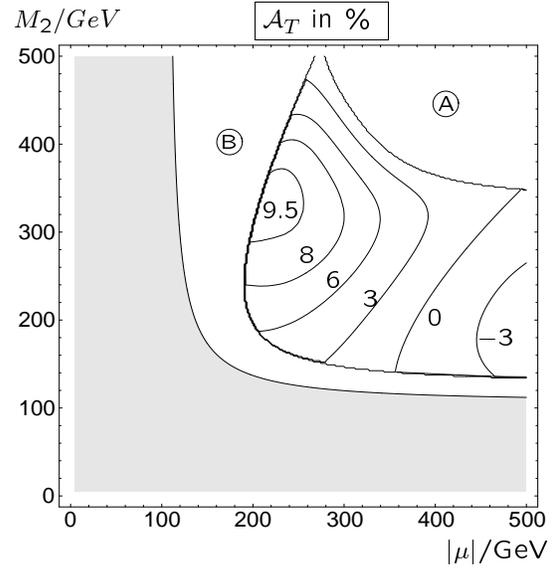}
}
\caption{
        Contour lines of the asymmetry ${\mathcal A}_{\rm T}$
        in the $|\mu|$--$M_2$ plane for $\varphi_{M_1}=0.5\pi $,
        $\varphi_{\mu}=0$, taking  $\tan \beta=10$, $m_0=100$ GeV,
        $\sqrt{s}=500$ GeV and $(P_{e^-},P_{e^+})=(0.8,-0.6)$.
        The area A (B) is kinematically forbidden since
        $m_{\tilde\chi^0_1}+m_{\tilde\chi^0_2}>\sqrt{s}$
        $(m_{\tilde\ell_R}>m_{\tilde\chi^0_2})$.
        The gray  area is excluded since
        $m_{\tilde\chi_1^{\pm}}<104 $ GeV \cite{olaf}.
        \label{plot_1}}
\end{figure}

In \cite{olaf} the dependence of  ${\mathcal A}_{\rm T}$ on
$\varphi_{M_1}$ and  $\varphi_{\mu}$ is analyzed.
In Fig.~\ref{plot_1} ${\mathcal A}_{\rm T}$ 
is shown in the $|\mu|$--$M_2$ plane for
$e^+e^-\to\tilde\chi^0_1 \tilde\chi^0_2$
with the subsequent decay of $ \tilde\chi^0_2$
into the right selectron and right smuon,
$\tilde \chi^0_2\to\tilde\ell_R\ell_1$.
${\mathcal A}_{\rm T}$
reaches values up to 9\%. Choosing  $\varphi_{M_1}=0.1\pi $
and $\varphi_{\mu}=0$, ${\mathcal A}_{\rm T}$ would still reach
values up to 5\%. A small value of $\varphi_{M_1}$ and in particular of
$\varphi_{\mu}$ is suggested by constraints on
electron and neutron electric dipole moments
for a typical SUSY scale of the order of a few 100 GeV.
The cross section 
$\sigma=\sigma(e^+e^-\to\tilde\chi^0_1\tilde\chi^0_2) \times 
{\rm BR}(\tilde \chi^0_2\to\tilde\ell_R\ell_1)\times
{\rm BR}(\tilde\ell_R\to\tilde\chi^0_1\ell_2)$
which is not shown, reaches values up to 60 fb.
Both  ${\mathcal A}_{\rm T}$ and
$\sigma$ also depend sensitively on the polarizations of the
$e^+$ and $e^-$-beams \cite{olaf}.

\section{Chargino and neutralino mass parameters at one-loop level}
\label{sec:loop}

As mentioned in Sec.~\ref{sec:pol} the neutralino
($\tilde{\chi}^0$) and chargino ($\tilde{\chi}^\pm$) mass
parameters can be extracted at lowest order from the masses and
production cross 
sections in $e^+e^-$ collisions with polarized beams \cite{CKMZ}.
At higher order, this extraction is not trivial and depends on
the renormalization scheme. In the scale dependent $\overline{\rm DR}$
scheme the one-loop corrections to the $\tilde{\chi}^0$ and $\tilde{\chi}^\pm$
mass matrices were calculated in \cite{piercelahanas}. For the
on-shell renormalization various methods were proposed
\cite{chmasscorr,fritzscheguasch}. They differ by different
counter terms for the parameters $M_1$, $M_2$ and $\mu$. Although the schemes
are equivalent in the sense that the observables (masses, cross
sections, etc.) are the same, 
the meaning of the parameters are different.

\begin{figure}
 \centering
 \mbox{\resizebox{75mm}{!}{\includegraphics{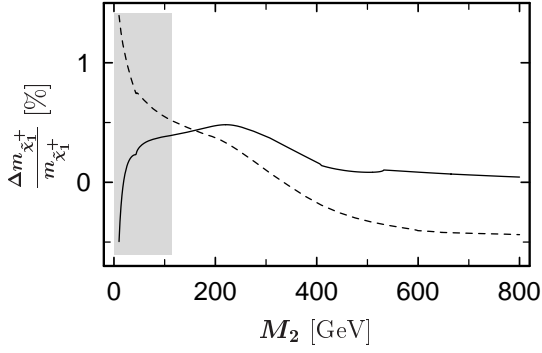}}}
 \caption[fig1]
 {Relative corrections to the
 ${\tilde\chi^+}_1$ mass,
 fixing $M_2$ and $\mu$ in the chargino (full lines) and
 neutralino (dashed lines) sector.
 The parameters are
 \{$m_{A^0}$, $\tan \beta$, $M_{\tilde Q_1}$, $M_{\tilde Q}$, $A$, $\mu$\} =
 \{500, 40/GeV, 300, 300, $-400$, $-220$\} GeV.
 The grey areas are excluded by the bound 
 $m_{\tilde\chi^+_1} \geq 100$ GeV \cite{fulloneloop}.}
 \label{fig:fixing1}
\end{figure}

\begin{figure}
 \centering
 \mbox{\resizebox{80mm}{!}{\includegraphics{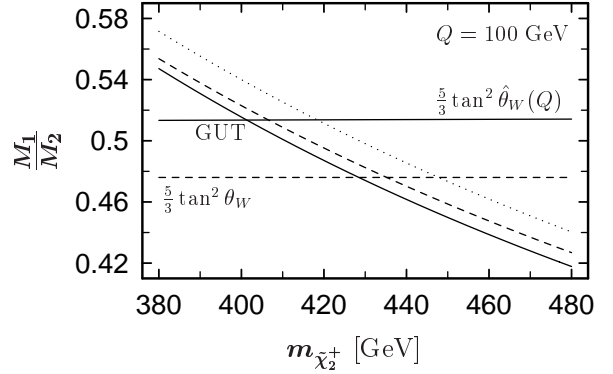}}}
 \caption[fig8]
 {The ratio $M_1/M_2$ as a function of the ${\tilde\chi^+}_2$ mass.
 The full, dashed and dotted line corresponds to the $\overline{\rm DR}$,
 on-shell \cite{chmasscorr}
 and effective \cite{fritzscheguasch} parameters.
 \{$m_{{\tilde\chi^+}_1}$, $m_{{\tilde\chi^0}_1}$, $\tan \beta$, $m_{A^0}$, $M_{\tilde Q_1}$, $M_{\tilde Q}$, $A$\} =
 \{135, 120, 20/GeV, 600, 350, 350, 500\} GeV \cite{fulloneloop}.}
 \label{fig:guttest}
\end{figure}

Using the scheme in \cite{chmasscorr}, it is shown in \cite{fulloneloop}
that at one-loop level 
the values of the on-shell parameters $M_2$ and $\mu$ depend on
whether they are determined from the $\tilde{\chi}^0$ or
$\tilde{\chi}^\pm$ system (see Fig. \ref{fig:fixing1}), while the
$\overline{\rm DR}$ parameters 
are equal in both sectors.
Assuming the SU(5) GUT relation for the $\overline{\rm DR}$
gaugino mass parameters, we obtain a finite shift for the on-shell values
$M_1=\frac{5}{3}\tan^2\theta_W\,M_2 + \Delta Y_{11}$.
In such a way it is possible to test the GUT relation (see
Fig. \ref{fig:guttest}).

\begin{acknowledgement}
This work was supported by the EU Network Programmes
HPRN-CT-2000-00148 and HPRN-CT-2000-00149,
by the \linebreak
''Fonds zur F\"orderung der wissenschaftlichen Forschung'' of
Austria, FWF Projects No.~P13139-PHY and No.~P16592-N02 and
by Spanish grant BFM2002-00345.
O.K. was supported by the 'Deutsche For\-schungs\-gemeinschaft'
(DFG) under contract Fr 1064/5-1 and by the EU
Research Training Site contract HPMT-2000-00124.
\end{acknowledgement}

\end{document}